# Web-enabling Cache Daemon for Complex Data

Ivan Voras, Mario Žagar
*University of Zagreb Faculty of Electrical Engineering and Computing*
*{ivan.voras, mario.zagar}@fer.hr*

**Abstract**. *One of the most common basic techniques for improving the performance of web applications is caching frequently accessed data in fast data stores, colloquially known as cache daemons. In this paper we present a cache daemon suitable for storing complex data while maintaining fine-grained control over data storage, retrieval and expiry. Data manipulation in this cache daemon is performed via standard SQL statements so we call it SQLcached. It is a practical, usable solution already implemented in several large web sites.*

## 1. Introduction

A quick survey (which does not attempt to be comprehensive) of Internet's most popular "generic" web applications and high-volume dynamic web sites confirms that most of them rely extensively on data cache daemons to help them achieve their high performance[1]. The results of this survey (from October 2007) are:

| Web site | Cache engine used |
|---|---|
| Slashdot (http://slashdot.org) | memcached [1] |
| Wikipedia (http://wikipedia.org) | memcached [1] |
| LiveJournal (http://livejournal.com) | memcached [1] |
| SourceForge (http://sourceforge.net) | memcached [1] |
| Google (http://google.com) | BigTable [2] |
| YouTube (http://youtube.com) | BigTable [2] |

**Table 1. Survey of cache engine usage on large Internet web sites**

We also observe that all of these web sites except Google use some of the rapid web application development languages and frameworks such as PHP, Python or Ruby, and have begun to rely on advanced caching techniques to maximize their performance. The popularity of *Memcached* in this survey, used by most large web sites without a major corporate backing, can be easily explained by the fact that it was the first Open Source cache daemon to provide generalized and consistent interfaces to most popular programming languages.

Out of a need to to increase performance in a complex web application developed at our Faculty[2], we've first created a cache layer based on Memcached, with which we've observed significant performance improvements. However, during the implementation and usage we have found that many common operations are not performed efficiently. These operations include: complex conditional data retrieval, complex cache expiry rules, and reduced need for serializing and unserializing data to and from strings. The lack of flexibility in the implemented solution (which only offers a simple key-value database) has lead us to consider a different approach. Finally, a project was started to implement a new cache daemon which can provide these features in an uniform and consistent way, with server-side implementation of most of the complex rules. The result is the *SQLcached*, whose architecture and implementation we present in this paper.

## 2. Data caching in currently common web applications

The basic idea behind using cache daemons in web applications is to skip repetitive CPU- and IPC- intensive steps by generating data only once and then storing it in a high-performance cache store, from where it can be retrieved as needed. A very common application of this idea is skipping the repetitive execution of complex or large SQL queries by integrating a cache layer between the web application and the database interface (library). This cache layer will commonly check if the result of the passed SQL query exists in the cache and if it does, it will return the data directly from the cache instead of passing the query for execution to the database. In this simple form, the simplest kind of directly addressable data store is enough to satisfy the required functionality.

---

[1]Note that this is different from generalized "HTTP cache" and "web acceleration" applications which cache resulting HTML and other content and act as a "black box" between the web server and its end users.

[2]The "Quilt" web CMS, implemented at the Faculty of Electrical Engineering and Computing, other University faculties and several government agencies in Croatia, made with the PHP language and using PostgreSQL database.

Common web cache daemons are essentially memory databases that offer a simple interface for storing and retrieving key-value records, with some "bonus" features like simple arithmetic operations (increment and decrement) for well-formed numeric values in the cache, simple atomic operations (get- and set-if-not-changed) and data expiry (an optional expiry timestamp attached to each key-value pair). A typical cache layer in web applications builds a key-value pair by hashing SQL query strings to form the key string and serializing database results to form the value string. Since the cache daemon often implements data expiry directly, it is enough for the application to check if the hashed SQL string exists in the cache and then either return the cached result (if it exists), or execute the query, store the result in the cache, and then return it.

## 2.1. Design of *Memcached*

One of the most common cache daemons in use on the web is *Memcached*, whose design is classical and straightforward. Since it's influenced the creation of *SQLcached*, we will present some of its major features here.

The architecture of *Memcached* makes it very efficient in the common case [3]. The cache daemon is implemented as a single executable that is resident (as its name implies) as a daemon process on a network-enabled server. It is created in C for POSIX environments and is most commonly deployed on Linux and other Unix-like operating systems. Applications communicate with the cache daemon using a simple text-based protocol (modelled after early TCP protocols like SMTP) which is implemented over TCP or UDP, with the TCP version being preferable. This protocol is easily implemented in practically any programming language, which has greatly helped *Memcached*'s popularity. The cache daemon uses a low-latency, asynchronous method of acquiring and handling network connections. The data store (key-value pairs) is organized as a dynamically sized hash table structure optimized for fast reading. The hash table size can only be increased (the condition for this is when the number of items in any hash bucket gets larger than two thirds of the base-two logarithm of the number of buckets). Memory for the hash table items is allocated using an internal slab allocator [4] whose purpose is to reduce memory fragmentation [5].

*Memcached* is a single-threaded daemon which handles network requests sequentially (at any given time, the cache daemon is actively working only on one request). This makes for a simple and efficient implementation, at the expense of not allowing it to scale on multi-CPU systems. However, because the execution of individual requests is very fast and the operating system handles network data transmission asynchronously while requests are being executed, the overall performance of *Memcached* is often more than sufficient for its purpose.

## 2.2. Shortcomings of *Memcached* and the design of *SQLcached*

*Memcached* is a proven solution for data caching and memory databases used by many large products, many of which are of "mission critical" importance for their respective companies. Most of its shortcomings are worked around in its implementation at the data cache layer in applications, though at the expense of efficiency and speed of operation. The purpose of this paper is not to discourage its use, but to point out some additional features that are generally useful for applications but which are not addressed in *Memcached* and to provide an alternative solution that does implement them. These features are:

*1. The ability to store complex data without excessive serialization.* While some forms of data conversion is always necessary to translate the data representation from the one used in the web application (e.g. in PHP) to the one used in the cache daemon, serialization in dynamic languages can be slow. We have observed that in PHP, the *serialize()* function is 1.5 times slower than a naïve approach when converting simple integer values, and up to 20 times slower when converting simple structures. Our analysis shows that much of this difference results from taking advantage of the programmer's knowledge of data types and structure layouts in the second case, versus the generality of the serialize() call.

This approach cannot be used with a simple key-value database.

*2. The ability to retrieve data sets based on complex criteria.* Many cache daemons, including *Memcached*, offer some way of retrieving a list of cached data records based on a provided list of keys, but retrieving keys based on a complex criteria such as "wildcard" matching of key strings, retrieving all cache items pertinent to a certain web page ID, etc. is not supported. Due to the inherent nature of the storage structure, hash tables (while very efficient at key-value pairs) cannot support many of the complex queries.

*3. The ability to expire data sets based on complex criteria.* Without additional metadata support, cache record expiry can commonly be performed either per-record (for each record individually) or in bulk (all records at once). Fine grained expiry would allow expiry only of certain data, for example all data pertinent to a web page ID.

*4. The ability to do complex operations on the data (both data-processing and algebraic).* This is a non-critical ability and is in any case more a convenience than a necessity. With a complex data model it should be possible to, for example, extend the expiry time ("time to live") of cached data items, or update only certain portions of the data, further increasing the cache efficiency when there are a lot of cached records or when the data generation is particularly slow.

Many of these features stem from the fact that most modern dynamic web pages are, as a rule, constructed of several individual elements. A simple example is a web page consisting of a header, a footer, a site navigation bar and the contents, each of which may have complex relationships with other elements on this and other pages. For example, the page's header and navigation bar can be common for all pages, while the content is changed by the administrators, and the footer contains the timestamp of the last change in the content. Adding a new page to the site means only the navigation bar is changed (for all pages). To support modern interactive sites (let's call them "Web 2.0" sites, for a lack of a better common name), individual users might have customized views of the same content, multiplying most of the cacheable elements per the number of users.

The lack of fine-grained control over data retrieval and expiry results in inefficient use of the cache. This inefficiency can be manifested by either of the two following scenarios. First, keeping too much logic and data in the application to avoid retrieving or expiring too many records is slow if the application is written in an interpreted language, and can result in many IPC calls to the cache to gather all needed data from the cache. Second, by considering the data at a too coarse granularity, keeping it in few large structures which are generated and transferred to and from the cache in bulk or expiring all data from the cache when a critical piece of user-visible data changes (versus expiring only the data pertinent to a certain user or to a certain web page), significant spikes in server load can be observed when new data is generated, which can sometime reduce the beneficial effects of having a cache. In our experience, this latter form of inefficiency is more dangerous to a smooth user experience (and a smooth and predictable server load).

*Memcached* lacks all of the enumerated features because it is structured as a true key-value database, with both the key and the value being simple opaque binary data strings. Some of the features can be emulated to an extent by folding data qualifiers into key strings, which was used by an early version of our web application. Our experience from the implementation was that such substitute techniques are often difficult to implement and negatively impact the overall performance of the system, as compared to what it could have been without them. This experience has motivated us to seek a different solution for the data cache.

Several different approaches were considered for the improvement of the overall system, among which are: modification of *Memcached* to better suit our data model, creation of a similar cache daemon which allows data to be properly separated into multiple independent domains (that can be expired separately), and creation of a complex caching daemon with the features of generic tabular database.

After weighing the benefits that each approach could bring to the system and the complexities of each implementation, the third option was chosen. Adding special support for our needs to *Memcached* would limit its usability when the needs change, and while separating keys into independent domains can help with expiry, retaining any form of key-value database doesn't do help with storage of complex data.

After the decision to implement a tabular memory database, we have considered an interface to the cache daemon. In the spirit of making it simple, a text interface was chosen. It was soon obvious that the most convenient (and easiest to learn) interface to the cache engine is a subset of SQL. In order not to duplicate work and create yet another SQL database, it was decided to base the new cache engine on SQLite [6].

## 3. Implementation of SQLcached

The new cache engine, which we named *SQLcached*, uses SQLite as its underlying storage engine. SQLite is a small, embeddable, serverless relational database with an interface that makes it usable for seamless integration into larger applications. At the time of this project's creation, two SQLite versions were in widespread use: version 2.8.17, a stable and mostly obsolete version, and version 3.3.5, with a newer architecture. Preliminary tests showed a significant difference in performance between the versions [7] in favour of the older version. Because of this, we have used version 2.8.17.

*SQLcached* is written in C and is designed as a daemon process for POSIX environments. Applications communicate with it via either the TCP or the "Unix sockets" interface (both can be used at the same time), using a simple text-based protocol. Network operations (reading, writing and connection handling) are implemented in an asynchronous way so that multiple simultaneous connections are handled efficiently, but only one request can be processed at the same time. The asynchronous requests are handled using the standard POSIX *poll()* interface. *SQLcached* can be deployed on more than one server to create a load-balancing setup.

The network protocol used by *SQLcached* offers the clients an almost complete set of SQL statements supported by SQLite, including relational features like n-way joins (though they may not be directly usable in a cache daemon, for performance reasons).

The SQLite backend is used as a memory-only database, and no data is ever committed to permanent storage. Indices can be created on appropriate database fields to speed up data retrieval. Both data and index structures are internally stored in dynamically balanced B-trees.

## 4. SQLcached features

The following sections describe how *SQLcached* implements features enumerated in section 2.2.

### 4.1. Support for storing arbitrary data without excessive serialization

*Sqlcached* stores data in tabular form, and the SQL interface exposed to the clients allows them to create arbitrary tables for their use. This feature allows applications to store entire structures in the cache without forced serialization, and later selectively retrieve only needed parts of the structures. The main benefit of this mode of operation is a smaller overhead in processing time and in memory allocation, for both the application and the cache daemon.

### 4.2. Support for retrieving data sets based on complex criteria

The flexibility of SQL allows applications using *SQLcached* to store and retrieve data using complex criteria. The ability to cache data per-user, per-page or per-application (with the same keys in each of the domains) can have a huge influence on efficiency of the cache layer, which doesn't have to fold all the data into a single namespace and can issue relatively complex queries to store and extract only the needed parts of complex structures. It also means that the usage of *SQLcached* is not limited to caching of opaque objects such as entire results of database queries, but allows a different approach in which applications can now cache their internal, already processed data. In addition to increasing efficiency and flexibility, this approach opens new possibilities for applications. *SQLcached* can be used for operations that would normally overtask the main database due to the high volume of data writes they generate, such as tracking of individual users' statistical information and presence (current web page or object they are interacting with).

### 4.3. Support for data expiry based on complex expiry rules

Data expiry in *SQLcached* can be performed either automatically or initiated by applications. Automatic data expiry can be triggered by one of three conditions, configurable by the applications: data age, number of records in the table and number of cache operations. Using these conditions, the data cache can be effectively limited in size and remain fresh enough for applications to use.

Applications that need more complex expiry can issue appropriate SQL commands with arbitrarily complex conditions, which can (for example) delete data cached per-user, per-page or per-application.

### 4.4. Support for complex operations on cached data

Applications using *SQLcached* have access to operators and functions supported by SQLite, which include basic arithmetic and comparison operators, simple numerical functions (e.g. ABS), data manipulation functions (e.g. UPPER) and aggregate functions (e.g. MIN, MAX).

## 5. Usability and performance of SQLcached

The proper application of *SQLcached* in web application requires a step away from the traditional design where the cache is used as a single-purpose key-value database. Instead, we propose a design where the applications exhibit more fine-grained control over the data and cooperate with the cache daemon to make the best use out of the rich set of features it offers. Using *SQLcached* as a simple key-value database is suboptimal, as shown on Figure 1, presenting results of benchmarking *SQLcached* and *Memcached* in a situation where the data records are simple key-value pairs with value sizes conforming to a geometric distribution[3]. In case of *SQLcached*, a single table was created with three columns: key, value, and time. Analogous write benchmarks follow the same trends and relations between the results.

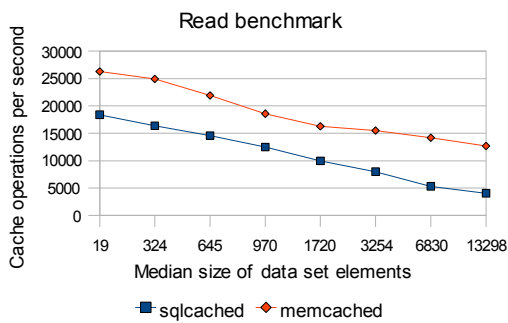

**Figure 1. Results of simple key-value read benchmarks.**

However, *SQLcached* offers features not present in *Memcached*, such as fine-grained data expiry. We have performed benchmarks on a data set of 100,000 records in 30,000 pages and 1000 users simulating cached data according to the example web site described in section 2.2[4] (appropriate indexes were created in the cache database schema). Table 2 contains results of forced data expiry benchmarks. Forced data expiry is frequently used when new content is posted which immediately obsoletes previously cached data (because users want to immediately see the effects of their actions).

| Cache daemon operation | Time |
| --- | --- |
| Memcached (expire entire set at once) | 1000.0 ms |
| SQLcached (expire cached data, a single page) | 0.2 ms |
| SQLcached (expire cached data, a single user) | 6.1 ms |

**Table 2. Effectiveness of fine-grained forced data expiry**

By employing fine-grained expiry of the cached data in the web CMS application used at our Faculty, we have observed up to 30% improvement in overall performance at periods of intensive content creation and significant reduction of load spikes in IO and CPU on the server, resulting in a more smooth and predictable operation. We believe that offloading more data operations from the cache layer in the application (written in PHP) to *SQLcached* could yield even better results.

## 6. Conclusion

This paper presented the *SQLcached* cache daemon, a network-enabled memory database intended to be used for caching often generated data in web applications. Development of *SQLcached* was motivated by the desire to increase performance in a real, production web application in a way that could not be done with the most popular open-source cache daemon, *memcached*.

*SQLcached* offers a highly flexible interface to client applications, based on a subset of SQL. This allows applications to perform complex operations on the cached data such as storage, retrieval and data expiry using complex rules, which results in a more efficient use of the cache. To gain maximum benefits from the advanced features of the cache database, applications using it must be modified not to treat cached data as simple binary strings, but as complex objects with arbitrary properties which can be used to increase efficiency. *SQLcached* is successfully implemented in our web application where it has satisfied all our requirements for a flexible cache database.

---

[3] These benchmarks were performed on an Athlon 64 2 GHz in 32-bit mode on FreeBSD 7 under directly comparable conditions.

[4] These benchmarks were performed on a Pentium M 1.5 GHz on FreeBSD 7 under directly comparable conditions.

The source code of *SQLcached* is published as Open source and is available at http://www.sf.net/projects/sqlcached.

## 7. Acknowledgements

This work is supported in part by the Croatian Ministry of Science, Education and Sports, under the research project "Software Engineering in Ubiquitous Computing".